\documentclass{article}
\usepackage{amsmath,amssymb,amsfonts}
\usepackage[margin=1in]{geometry}
\usepackage{tikz}
\def\Point#1#2#3{\node [fill=black,inner sep=1pt,label=#1:{#2}] at #3 {}}
\def\Label#1#2#3{\node [label=#1:{#2}] at #3 {}}
\newcount\pno \def\mn{\medskip\noindent}
\def\makenote#1#2{\expandafter\def\csname#1\endcsname{\advance\pno by 1
\medskip\noindent\textbf{#2 \the\pno. }}}\makenote{defn}{Definition}
\makenote{ex}{Example}\makenote{thm}{Theorem}\makenote{lem}{Lemma}
\makenote{rem}{Remark}\makenote{prop}{Proposition}\makenote{cor}{Corollary}
\def\beginalgo{\begingroup\newcount\lineno\global\advance\lineno by 1%
\noindent{\small\the\lineno\quad}%
\def\n{\\{\global\advance\lineno by 1\small\the\lineno\quad\normalsize}}%
\def\_{\qquad}\def\b##1{{\bf ##1}}\def\={\leftarrow}\normalsize}
\def\endalgo{\endgroup}
\def\qed{\hfill$\square$}\def\mn{\medskip\noindent}\def\imp{\Rightarrow}
\def\ts{\textstyle}
\def\A{\mathcal A}\def\d{\mathrm d}\def\E{\mathbb E}\def\P{\mathbb P}

\def\Cov{\mathrm{Cov}}\def\Var{\mathrm{Var}}
\begin{document}
\begin{center}{\Large\textbf{Note on the capacity
and geometric realizability\\of combinatorial mobile sensor networks}}\\
Yizhen Chen

\bigskip\textbf{Abstract}

\medskip\begin{minipage}{.75\textwidth}
We develop the mathematical theory of a model, constructed by
C.\ Gu, I.\ Downes, O.\ Gnawali, and L.\ Guibas, of networks that
diffuse continuously acquired information from mobile sensor nodes.
We improve estimates of the expectation and variance of capacity
of their model of \textit{restricted combinatorial mobile sensor networks}
(RCMSN) and \emph{geometric mobile sensor networks} (GMSN),
and give the maximum capacity of a variant of GMSN.
We also show that the problem of deciding when an RCMSN is generated
from a GMSN is NP-Hard, while a simple variant is solvable in polynomial
time.\end{minipage}

\bigskip\bf1. Introduction\end{center}

A \emph{combinatorial mobile sensor network} (CMSN) models
$n$ sensors \emph{sensors} that continuously receive and and store
information. At a \emph{communication event} of two sensors,
they receive additionally all information the other has stored.
In order to evaluate the capacity of information diffusions in a CMSN,
[GDGG] collects into an \emph{information packet} all information received
by a sensor between two successive communication events, and merges
the information packets from the two sensors at each communication event,
as after the event all the information in the packet always goes together.
Thus, if there are $k$ communcation events in total, then $k$
information packets are generated. In an ideal situation, all sensors receive
all packets eventually. The capacity is measured by comparing with this.

\defn A \textbf{combinatorial mobile sensor network} (CMSN) 
of $n$ sensors, numbered by $1,2,\ldots,n$, is a sequence of
\emph{packets} $a_1,\ldots,a_N\in\{\{x,y\}:1\le x<y\le n\}$.
A \textbf{restricted combinatorial mobile sensor network} (RCMSN)
of $n$ sensors is a CMSN where $N=\binom n2$ and $a_1,\ldots,a_N$ are distinct.

A packet $a_k$ \textbf{reaches} sensor $x$, or that there is a
\textbf{delivery} from $a_k$ to $x$, if there exist $k=k_0<k_1<\cdots<k_m$
such that $x\in a_{k_m}$ and $a_{k_j}\cap a_{k_{j-1}}\ne\emptyset$ for
each $1\le j\le m$. We say $a_k$ reaches $x$ \textbf{in $m$ hops} if
$m\ge0$ is the minimum number satisfying the condition.

The \textbf{capacity} of a CMSN is
$\frac1{nN}|\{(k,x):a_k\text{ reaches }x\}|$.

\medskip [GDGG] also studies a specific type of CMSN where the sensors
move with constant speed on lines in the plane from left to right
and communicate whenever a pair meet.

\defn A \textbf{geometric mobile sensor network} (GMSN) of $n$ sensors
is an RCMSN of $n$ sensors such that there are non-vertical
lines $l_1,\ldots,l_n$ with distinct slopes, and the RCMSN is given
by ordering pairs $\{i,j\}$ according to the $x$-coordinate of $l_i\cap l_j$.

A \textbf{restricted geometric mobile sensor network} (RGMSN) of $n$ sensors
and $k$ slopes is a CMSN of $n$ sensors such that there are non-vertical
lines $l_1,\ldots,l_n$ with $\le k$ distinct slopes, and the CMSN is
given by ordering pairs $\{i,j\}$ according to the $x$-coordinate of
$l_i\cap l_j$.

\medskip The authors of [GDGG] claimed that the capacity of an RCMSN
of $n$ sensors is $1-O(\frac{\log^2n}n)$ with probability
$\ge1-O(\frac{\log n}n)$, and the variance of the capacity is
$O(\frac{\log^2n}n)$. They also showed that the expected capacity of
a GMSN when the slopes and intercepts of its lines are chosen independently
and uniformly in closed intervals is in $[\frac23,\frac56]$.
Geneson showed that the maximum and minimum capacities of an RCMSN are
$1-\frac1n+\frac2{n^2}$ and $\frac23(1+\frac1n)$ respectively, and
both are realizable as GMSN [G]. For a simpler proof of this maximum
capacity, see the Appendix.

We shall improve these results.

\thm If the packets of an RCMSN are chosen uniformly from permutations
of $\{\{x,y\}:1\le x<y\le n\}$, then the expected capacity is
$1-\frac{\log n+\gamma-1}n+O(\frac{\log^2n}{n^2})$. Thus the variance
of the capacity is $O(\frac{\log n}n)$.

\thm The expected capacity of a GMSN with lines $l_k:y=a_kx+b_k$
where $(a_k,b_k)$ are chosen independently and uniformly from
$[a^-,a^+]\times[b^-,b^+]$ is $\frac56+O(\frac1n)$, and the variance
of the capacity is $(\frac{493}{2160}-\frac{\pi^2}{45})\frac1n(1+o(1))$.

\thm The maximum capacity of an RGMSN of $n$ sensors and $k$ slopes
is $\frac12+\frac1n$ when $k=2$,
$1-\frac1{\sqrt n}+\frac9{8n}+O(\frac1{n\sqrt n})$ when $k=3$,
and $1-\frac{k-2}{(k-3)n}+O(\frac1{n^2})$ when $k>3$.

\medskip We also study whether a given RCMSN or CMSN is realizable as
a GMSN or an RGMSN with a given number of slopes.

\thm Deciding if a given RCMSN is realizable as a GMSN is NP-Hard.

\prop There is a polynomial-time algorithm deciding if a given CMSN is
realizable as an RGMSN of two slopes.

\medskip For some secondary and technical results concerning the exact
expected and maximum capacities of an RGMSN, see the previous version
of this paper (arXiv:1910.04162v3). The proof of Theorem 22
in that version is incorrect, and the statement is an open problem.

\begin{center}\bf2. Expected capacity of restricted mobile sensor networks
\end{center}

It is claimed in [GDGG, Thm.~3.2] that the capacity of an RCMSN of $n$ sensors
is $1-O(\frac{\log^2n}n)$ with probability $\ge1-O(\frac{\log n}n)$,
but Geneson found an error in their proof [G]. We shall correct the proof
and improve the estimate.

\mn\textit{Proof of Theorem 3.} We generate a uniform random permutation
of the packets $P=\{\{x,y\}:1\le x<y\le n\}$ by ordering $p=\{x,y\}$
according to an independent and uniformly random time $t_p\in[0,1]$.
We denote ``$p$ reaches $z$'' by $p\to z$. Then the expected capacity
is $\P(p\to z)$, where $p\in P$ and $z\in[n]$ are independent and uniformly
random. Since $\{1,2\}\to 1,2$, we have
$\P(p\to z)=\P(\{1,2\}\to z)=1-\frac{n-2}n\P(\{1,2\}\not\to3)$ by symmetry.

Conditioning on $t_{1,2}=t$, the other times $t_{x,y}$ are independent
with $\P(t_{x,y}>t)=1-t$. Conditioning further on $t_{x,y}>t$, the times
$t_{x,y}$ satisfying this are independent and uniformly random in $[t,1]$.
We may move the interval $[0,t]$ to $[1,1+t]$ and shift by $-t$:
let $t_3$ be the minimum $t_p$ such that $p\ni3$ is reached, assuming
$1,2$ are reached at time 0, then
$\P(\{1,2\}\not\to3\mid t_{1,2}=t)=\P(t_3>1-t)$,
so $\P(\{1,2\}\not\to3)=\int_0^1\P(t_3>1-t)\,\d t=\E[t_3]$.

Assuming $1,2$ are reached at time 0, at the first time $T_k\in[0,1]$ when
$k$ points are reached, let $R_k$ be the set of points reached, then
$R_2=\{1,2\}$ and $T_2=0$. Let $X_k=T_{k+1}-T_k$ be the waiting time,
and $\A_k$ the $\sigma$-algebra generated by all pairs $p\in P$ and
their $t_p$ when $t_p\le T_k$. [CRRZ, Lemma 4.3] showed that
\[ \frac{1-T_k}{k(n-k)+1}\le\E[X_k\mid\A_k]\le\frac1{k(n-k)+1}. \]
(Their result actually starts with $R_1=\{1\}$ and $T_1=0$ instead,
but the same proof applies to our case.)

We have $t_3=\sum_{k=2}^{n-1}X_k1_{3\notin R_k}$ and
$\P(3\notin R_k)=\frac{n-k}{n-2}$ by symmetry ($3\notin R_k$ iff 3 is
one of the $n-k$ points in $\{3,\ldots,n\}$ not in $R_k$). So
\begin{align*} \E[X_k1_{3\notin R_k}]&=\E[\E[X_k\mid\A_k]1_{3\notin R_k}]
\le\frac{\P(3\notin R_k)}{k(n-k)+1}=\frac{n-k}{(n-2)(k(n-k)+1)}\\
\imp\E[t_3]&=\sum_{k=2}^{n-1}\E[X_k1_{3\notin R_k}]\le
\sum_{k=2}^{n-1}\frac{n-k}{(n-2)(k(n-k)+1)}\\
&=\frac{H_{n-1}-1}{n-2}-\frac1{n-2}\sum_{k=2}^{n-1}\frac1{k(k(n-k)+1)}
=\frac{H_{n-1}-1}{n-2}+O(n^{-2}).\end{align*}We also have
\[\E[T_k]=\sum_{j=2}^{k-1}\E[X_j]\le\sum_{j=2}^{k-1}\frac1{j(n-j)+1},\]
so \begin{align*}
\E[t_3]&\ge\sum_{k=2}^{n-1}\frac{\P(3\notin R_k)-\E[T_k]}{k(n-k)+1}\\
&=\frac{H_{n-1}-1}{n-2}+O(n^{-2})-\sum_{k=2}^{n-1}\sum_{j=2}^{k-1}
\frac1{(k(n-k)+1)(j(n-j)+1)}\\
&\ge\frac{H_{n-1}-1}{n-2}+O(n^{-2})
-\frac12\bigg(\sum_{k=2}^{n-1}\frac1{k(n-k)+1}\bigg)^2=
\frac{H_{n-1}-1}{n-2}-O\bigg(\frac{\log^2n}{n^2}\bigg)
\end{align*}
which gives the result as the expected capacity is $1-\frac{n-2}n\E[t_3]$.
The variance result follows as the capacity is in $[0,1]$.\qed

\mn\textit{Proof of Theorem 4.} For each three distinct lines
$l_1,l_2,l_3$ with $a_1<a_2<a_3$, there are three possible deliveries
(from the intersection of two to the third), of which two deliveries
are guaranteed in one hop, and the other one happens in at most two hops
[GDGG, Thm. 3.4 and 3.8]. Thus in the figure below,\\in the left case
($\frac{b_3-b_2}{a_3-a_2}<\frac{b_3-b_1}{a_3-a_1}<\frac{b_2-b_1}{a_2-a_1}$),
$(2,3)\to1$ iff there is a line $l$ in the network with slope less than
that of $l_1$ such that $l\cap l_2$ is on the right of $l_3\cap l_2$
(in $P_L:=\{(a,b):a<a_1\text{ and }
\frac{b-b_2}{a-a_2}<\frac{b_3-b_2}{a_3-a_2}\}$). This only
happens in $C_L:=\{b_1<\frac{b_3-b_2}{a_3-a_2}(a_1-a_2)+b_2<b^+\}$;
\\in the right case
($\frac{b_3-b_2}{a_3-a_2}>\frac{b_3-b_1}{a_3-a_1}>\frac{b_2-b_1}{a_2-a_1}$),
$(1,2)\to3$ iff there is a line $l$ in the network with slope greater than
that of $l_3$ such that $l\cap l_2$ is on the right of $l_1\cap l_2$
(in $P_R:=\{(a,b):a>a_3\text{ and }
\frac{b-b_2}{a-a_2}<\frac{b_2-b_1}{a_2-a_1}\}$). This only happens
in $C_R=\{b^-<\frac{b_2-b_1}{a_2-a_1}(a_3-a_1)+b_1<b_3\}$.
\\In both cases we say that \textbf{the last delivery of $\{1,2,3\}$
happens via $l$}.

\begin{center}\begin{tikzpicture}[scale=0.5]
\draw(-3,0)--(3,0);\draw(-3,1)--(3,-2);
\draw(-1,-3)--(2,3);\draw[dashed](0,3)--(3,-3);
\Label{left}{$l_3$}{(-1,-3)};\Label{right}{$l_2$}{(3,0)};
\Label{right}{$l_1$}{(3,-2)};\Label{left}{$l$}{(0,3)};
\node [fill=black,inner sep=1pt] at (0.5,0) {};
\end{tikzpicture}\qquad\qquad\begin{tikzpicture}[scale=0.5]
\draw(-3,0)--(3,0);\draw(-3,-1)--(3,2);
\draw(-1,3)--(2,-3);\draw[dashed](0,-3)--(3,3);
\Label{left}{$l_1$}{(-1,3)};\Label{right}{$l_2$}{(3,0)};
\Label{right}{$l_3$}{(3,2)};\Label{left}{$l$}{(0,-3)};
\node [fill=black,inner sep=1pt] at (0.5,0) {};
\end{tikzpicture}\end{center}

The two cases are disjoint, so $C_L\cap C_R=\varnothing$.
In $C_L$, we have $|P_L|>0$, so the probability that one line is in $P_L$
approaches 1 as $n\to\infty$, and similar for $C_R$. Thus, the expected
capacity of a GMSN $\to\frac23+\frac13(\P(C_L)+\P(C_R))$ as $n\to\infty$,
where the first $\frac23$ is the guaranteed one-hop delivery.

By affine transformations, we may assume $a^-=b^-=0$ and
$a^+=b^+=1$. For fixed $a_1,a_2,a_3,b_1,b_3$, the sets
$C_L$ and $C_R$ for $b_2$ are intervals in $[0,1]$ of lengths
$\frac{a_3-a_2}{a_3-a_1}(1-b_1)$ and $\frac{a_2-a_1}{a_3-a_1}b_3$
respectively, so \begin{align}\nonumber\P(C_L\mid a_1,a_2,a_3)&=
\int_0^1\frac{a_3-a_2}{a_3-a_1}(1-b_1)\,\d b_1=\frac{a_3-a_2}{2(a_3-a_1)}\\
\text{and }\P(C_R\mid a_1,a_2,a_3)&=
\int_0^1\frac{a_2-a_1}{a_3-a_1}b_3\,\d b_3=\frac{a_2-a_1}{2(a_3-a_1)}.
\end{align} Hence \begin{equation}\P(C_L)+\P(C_R)=\frac12,\end{equation}
and the expected capacity of a GMSN $\to\frac56$ as $n\to\infty$.

Next, we compute the variance of the capacity. Let $X_{ijk}$ be
the indicator of all three deliveries of $l_i,l_j,l_k$ being reached,
then the desired variance is \begin{align*} \Var
&=\ts(n\binom n2)^{-2}\sum_{|I|=|J|=3}\Cov(X_I,X_J)\\
&=\ts(n\binom n2)^{-2}\big(\binom n3\binom{n-3}3\Cov(X_{123},X_{456})\\
&+\ts n\binom{n-1}2\binom{n-3}2\Cov(X_{123},X_{145})+O(n^4)\big)\\
&=\ts(\frac19\Cov(X_{123},X_{456})+\frac1n\Cov(X_{123},X_{145})+\frac4{n^2})
(1+o(1))\end{align*} So it remains to compute the two covariances
in this formula.

\mn\textit{Claim.} $\Cov(X_{123},X_{456})=o(\frac1n)$.

Let $X_{123}'$ and $X_{456}'$ be the indicators that the last delivery
of $\{1,2,3\}$ and $\{4,5,6\}$ are reached via some $l_i$ ($7\le i\le n$),
respectively, and let $P_{123}$, $P_{456}$ be their corresponding
$P_L$ or $P_R$, and $C_{123}$, $C_{456}$ their corresponding $C_L$ or $C_R$
(without assuming an order $a_1<a_2<a_3$).
The probability that some $l_i$ is in $P_{123}$ is
$\E[X_{123}']=1-(1-|P_{123}|)^{n-6}$, and the probability that some $l_i$
is in $P_{123}$ and some $l_j$ is in $P_{456}$ is \[\E[X_{123}'X_{456}']
=1-(1-|P_{123}|)^{n-6}-(1-|P_{456}|)^{n-6}+(1-|P_{123}\cup P_{456}|)^{n-6}.\]
By independence, the covariance is 0 when $|P_{123}||P_{456}|=0$, so
\begin{equation}\Cov(X_{123}',X_{456}')=\E[1_{|P_{123}|\ne0\ne|P_{456}|}
(1-|P_{123}\cup P_{456}|)^{n-6}]-\E[1_{|P_{123}|\ne0}(1-|P_{123}|)^{n-6}]^2.
\end{equation} We have \begin{align*}0
&\le(1-|P_{123}|)^{n-6}(1-|P_{456}|)^{n-6}-(1-|P_{123}|-|P_{456}|)^{n-6}\\
&\le(n-6)|P_{123}||P_{456}|(1-|P_{123}|)^{n-7}(1-|P_{456}|)^{n-7}\\
&\le(n-6)|P_{123}||P_{456}|e^{(7-n)(|P_{123}|+|P_{456}|)}\\
&\le\tfrac{n-6}{(n-7)^2}e^{(7-n)(|P_{123}|+|P_{456}|)/2},\end{align*}
and when $P_{123}\cap P_{456}\ne\varnothing$, \begin{align*}0
&\le(1-|P_{123}\cup P_{456}|)^{n-6}-(1-|P_{123}|-|P_{456}|)^{n-6}\\
&\le(1-\tfrac12(|P_{123}|+|P_{456}|))^{n-6}-(1-|P_{123}|-|P_{456}|)^{n-6}\\
&\le\tfrac{n-6}2(|P_{123}|+|P_{456}|)(1-\tfrac12(|P_{123}|+|P_{456}|))^{n-7}\\
&\le\tfrac{2(n-6)}{n-7}e^{(7-n)(|P_{123}|+|P_{456}|)/4}
\end{align*}
We also have $(1-|P_{123}|)^{n-6}\le e^{(6-n)|P_{123}|}$ and \begin{align*}
\P(X_{123}\ne X_{123}')&\le\E[(1-|P_{123}|)^{n-6}(1-(1-|P_{123}|)^3)]\\
&\le3\E[|P_{123}|e^{(6-n)|P_{123}|}]\le\tfrac3{n-6}\E[e^{(6-n)|P_{123}|/2}],
\end{align*} so by (3), to prove the claim, it remains to estimate
\begin{align}\E[1_{|P_{123}|\ne0}e^{-\lambda|P_{123}|}]&=o(1)\text{ and}\\
\E[1_{|P_{123}|\ne0\ne|P_{456}|,P_{123}\cap P_{456}\ne\varnothing}
e^{-\lambda(|P_{123}|+|P_{456}|)}]&=o(\lambda^{-1}).\end{align}

First, we estimate $\E[1_{|P_{123}|\ne0}e^{-\lambda|P_{123}|}]$.
Without loss of generality, by transformation
$(y=ax+b)\mapsto(y=(1-a)x+(1-b))$, the lines $l_1,l_2,l_3$ are in $C_R$ and
$P_{123}=P_R$.

The region $P_R$ is triangular if the line connecting
$(a_1,b_1)$ and $(a_2,b_2)$ has negative slope $k=-\frac{b_2-b_1}{a_2-a_1}>0$,
and intersects with the segment $[0,1]\times\{0\}$. In this case, $P_R$
is a triangle with height $h:=-k(a_3-a_1)+b_1\le k(1-a_3)$ and width
$h/k$ (see the figure below). The $e^{-\lambda|P_R|}$-weighted probability
density that $P_R$ is triangular with $a_3\in[a,a+da]$ is (note that
$db_2\,db_1=(a_2-a_1)\,dk\,dh$) \begin{align} \rho_\lambda^R(a)&=
6\int_0^a\!\!\int_{a_1}^a\!\int_0^1\!\!\int_0^\infty\!\!\!
\int_h^1e^{-\lambda h^2/(2k)}1_{h\le k(1-a),b_1=h+k(a-a_1)\le1}(a_2-a_1)
\,\d b_3\,\d k\,\d h\,\d a_2\,\d a_1\nonumber\\
&\lesssim\int_0^a\!\!\int_{a_1}^a\!\int_0^{1/(a-a_1)}\!\!\int_0^\infty
e^{-\lambda h^2/(2k)}(a_2-a_1)\,\d h\,\d k\,\d a_2\,\d a_1\sim\lambda^{-1/2}.
\end{align} Similarly, the
$\frac hke^{-\lambda|P_R|}$-weighted probability density that
$P_R$ is triangular with $a_3\in[a,a+da]$ is \begin{equation}
w_\lambda^R(a)\lesssim\lambda^{-1}.\end{equation} The corresponding densities
$\rho_\lambda^L(a)=\rho_\lambda^R(1-a)$ and $w_\lambda^L(a)=w_\lambda^R(1-a)$
have the same estimates for the region $P_L$.

\begin{center}\begin{tikzpicture}[scale=0.5]
\draw(0,0)--(0,5)--(5,5)--(5,0)--(0,0);\draw(1,3)--(2,2);
\draw[dashed](2,2)--(4,0);\draw[dashed](3,4)--(3,0);
\fill[color=red!20](3,1)--(4,0)--(3,0)--(3,1);
\Point{above}{$(a_1,b_1)$}{(1,3)};\Point{left}{$(a_2,b_2)$}{(2,2)};
\Point{above}{$(a_3,b_3)$}{(3,4)};\Label{left}{$h$}{(3,0.5)};
\Label{below}{$h/k$}{(3.5,0)};\Label{left}{1}{(0,5)};
\Label{below}{0}{(0,0)};\Label{left}{0}{(0,0)};\Label{below}{1}{(5,0)};
\Label{below}{triangular}{(2.5,-1)};
\end{tikzpicture}\qquad\qquad\begin{tikzpicture}[scale=0.5]
\draw(0,0)--(0,5)--(5,5)--(5,0)--(0,0);\draw(1,3)--(3,2);
\draw[dashed](3,2)--(5,1);\draw[dashed] (4,4)--(4,0);
\fill[color=red!20](4,0)--(4,1.5)--(5,1)--(5,0)--(4,0);
\Point{above}{$(a_1,b_1)$}{(1,3)};\Point{left}{$(a_2,b_2)$}{(3,2)};
\Point{above}{$(a_3,b_3)$}{(4,4)};\Label{left}{$h$}{(4,0.75)};
\Label{below}{$1-a_3$}{(4.5,0)};\Label{left}{1}{(0,5)};
\Label{below}{0}{(0,0)};\Label{left}{0}{(0,0)};
\Label{below}{non-triangular}{(2.5,-1)};
\end{tikzpicture}\end{center}

If $P_R$ is not triangular, then $|P_R|\ge\frac12h(1-a_3)$ (see the figure
above), so \begin{align*}\E[1_{|P_R|\ne0}e^{-\lambda|P_R|}
\mid P_R\text{ not triangular}]
&\lesssim\int_0^1\!\!\int_0^1\!\!\int_0^1\!\!\int_0^1\!\!\int_0^{1/(a_2-a_1)}
e^{-\lambda h(1-a_3)/2}(a_2-a_1)\,\d k\,\d h\,\d a_2\,\d a_1\,\d a_3\\
&\le\int_0^1\!\!\int_0^1e^{-\lambda ha_3/2}\,\d h\,\d a_3
\lesssim\lambda^{-1}\log\lambda.
\end{align*} Hence we have the first estimate (4):
$\E[1_{|P_R|\ne0}e^{-\lambda|P_R|}]=O(\lambda^{-1/2})$.

For the second estimate (5), since $P_{123}$ and $P_{456}$ are independent,
the expectation is $O(\lambda^{-1/2})\cdot O(\lambda^{-1}\log\lambda)=
O(\lambda^{-3/2}\log\lambda)$ by (6) and (7) if one of them is not
triangular, so it remains to estimate (5) when both are triangular.
Let $a,a'$ be the $x$-coordinates of the right-angle vertex of
the triangles, and $w,w'$ the widths of the triangles, then
$|a-a'|\le\max(w,w')$, so either $|a-a'|\le2w$ or $|a-a'|\le2w'$.
Thus the expectation (5) is
\[\le\int\!\int_{|a-a'|\le2w}\rho_\lambda(a)\,\d a'\,\d(a,w)\lesssim
\lambda^{-1/2}\int w\,\d(a,w)\lesssim\lambda^{-3/2}\]
by (6) and (7). This proves the claim.

Finally, we compute $\Cov(X_{123},X_{145})$. We have $X_{123}\to1_{C_{123}}$
as $n\to\infty$, and $X_{123},X_{145}$ are independent conditional on $l_1$,
so $\Cov(X_{123},X_{145})\to\Var(f(l_1))$ where $f(l_1):=\P(C_{123}\mid l_1)$.
By (2), $\E[f(l_1)]=\frac12$. Let $g(l_1)$ be the probability of
$C_{123}$ conditional on $l_1$, given that $C_{123}$ is $C_R$.

By (1), we have \begin{align*}
g(l_1)&=\int_{a_1}^1\int_{a_2}^1\frac{a_2-a_1}{2(a_3-a_1)}\,\d a_3\,\d a_2
=\frac{(a_1-1)^2}8\text{ if }a_1<a_2<a_3,\\
g(l_1)&=\int_0^{a_1}\int_{a_3}^{a_1}\frac{a_2-a_3}{a_1-a_3}b_1\,\d a_2\,\d a_3
=\frac{a_1^2b_1}4\text{ if }a_1>a_2>a_3,\text{ and}\\
g(l_1)&=\int_0^{a_1}\int_{a_1}^1\int_0^1
\phi\bigg(b_1+\frac{b_2-b_1}{a_2-a_1}(a_3-a_1)\bigg)\,\d b_2\,\d a_3\,\d a_2
\text{ if }a_2<a_1<a_3,\end{align*} where $\phi(x)=(1-x)1_{x\in[0,1]}$.
Let $h(a_1,b_1)$ be the last integral, then
\[ h(a_1,b_1)+h(1-a_1,1-b_1)=\int_0^{a_1}\int_{a_1}^1
\bigg(J_{b_1}\bigg(\frac{a_3-a_1}{a_1-a_2}\bigg)+
J_{1-b_1}\bigg(\frac{a_1-a_2}{a_3-a_1}\bigg)\bigg)\,\d a_3\,\d a_2,\]
where $J_b(t)=\int_0^1\phi(b+(b-s)t)\,\d s$ satisfies
$J_b(t)+J_{1-b}(t^{-1})=(1+t)\min(1-b,b/t)$, so
\begin{align*} h(a,b)+h(1-a,1-b)
&=\begin{cases}j(a,b)&a+b\le1\\j(1-a,1-b)&a+b>1\end{cases}\\
\text{where }j(a,b)&=ab-\frac{a^2b}{4(1-b)}+
\frac{a^2b}2\log\bigg(\frac{(1-a)(1-b)}{ab}\bigg).\end{align*}
Thus, by symmetry, \begin{align*} \E[f(l_1)^2]
&=\int_0^1\int_0^1(g(a,b)+g(1-a,1-b))^2\,\d b\,\d a
=2\int_0^1\int_0^{1-a}(g(a,b)+g(1-a,1-b))^2\,\d b\,\d a\\
&=2\int_0^1\int_0^{1-a}\bigg(\frac{a^2+(1-a)^2}4+\frac{a^2b+(1-a)^2(1-b)}2
+2j(a,b)\bigg)^2\,\d b\,\d a\\
&=\frac{1033}{2160}-\frac{\pi^2}{45},
\end{align*} so $\Cov(X_{123},X_{145})\to\Var(f(l_1))=
\E[f(l_1)^2]-\E[f(l_1)]^2=\frac{493}{2160}-\frac{\pi^2}{45}$.\qed

\begin{center}\bf3. Maximum capacity of restricted geometric
mobile sensor networks \end{center}

\noindent\textit{Proof of Theorem 5.} The $a$ lines with the largest
slope $M$ and the $b$ lines with the smallest slope $m$ form a grid.
The intersection of the $i$th line from the left with slope $M$ and
the $j$th line from the left with slope $m$ cannot reach at least $i+j-2$
lines with slope $M$ or $m$. In total, $\sum_{i=1}^a\sum_{j=1}^b(i+j-2)$
deliveries are not made.

The grid of these $a+b$ lines is the union of $a+b$ curves as the figure
below. The intersection between the $i$th curve from the left and
each of the remaining $n-a-b$ lines cannot reach $j-1$ lines in the grid,
resulting in $(n-a-b)\sum_{i=1}^{a+b}(i-1)$ deliveries not made in total.

\begin{center}\begin{tikzpicture}[scale=0.5]
\draw(3,1)--(2,0)--(3,-1);\draw[dashed](-3,1)--(-2,0)--(-3,-1);
\draw(-2,2)--(-1,1)--(-2,0)--(-1,-1)--(-2,-2);
\draw(1,3)--(0,2)--(1,1)--(0,0)--(1,-1)--(0,-2)--(1,-3);
\draw[dashed](2,2)--(1,1)--(2,0)--(1,-1)--(2,-2);
\draw[dashed](-1,3)--(0,2)--(-1,1)--(0,0)--(-1,-1)--(0,-2)--(-1,-3);
\end{tikzpicture}\end{center}

Also, the $i$th intersection from the right among all lines can reach
at most $i+1$ lines, so the $n-1$ intersections from the right cannot
reach at least $\sum_{i=1}^{n-2}i$ lines in total. At most $(n-1)(a+b)$
of these non-deliveries are counted in the previous two paragraphs.
Thus, the maximum capacity is at most \begin{align*}\kappa_0
&=1-\frac{(\sum_{i=1}^{n-2}i)+(n-a-b)(\sum_{k=1}^{a+b}(i-1))
+\sum_{i=1}^a\sum_{j=1}^b(i+j-2)-(n-1)(a+b)}{
n(ab+(a+b)(n-a-b)+\frac{(k-2)(k-3)}{2}(\frac{n-a-b}{k-2})^2)}\\
&=1-\frac{a+b-2}{2n}-\frac{\frac{1}{2}(n-a-b-1)(n-2)-\frac{1}{2}(a+b)^2
-\frac{1}{4}(1-\frac{1}{k-2})(a+b-2)(n-a-b)^2}{n(ab+(a+b)(n-a-b)
+\frac{k-3}{2(k-2)}(n-a-b)^2)}.\end{align*}
When $a+b<n$ is fixed, only the $ab$ on the denominator can vary,
so when $\kappa_0$ is minimized, either $a=b$
(when the numerator is positive) or $b=1$
(when the numerator is negative). In either case, the formula
of $\kappa_0$ contains only one free variable, so we find that when
$k\ge4$, the minimum of $\kappa_0$ is either
\[ \frac{(k-2)n^2-(5k-10)n+2k-4}{(k-3)n^2+4n-2k} \]
or a real root $t$ between 0 and $n$ of \[ (-(k-2)^3n^5+O(n^4))+
((k-2)^2(k-3)n^5+O(n^4))t+O(n^4)t^2+O(n^3)t^3+O(n^2)t^4=0. \]
Both result in an upper bound $1-\frac{k-2}{(k-3)n}+O(n^{-2})$
of the capacity.

Consider $n-2$ lines with $k-2$ different slopes, a line with the smallest
slope to the right of all their intersections, and a line $l$ with
the largest slope to the right of all intersections of the other lines.
Then all intersections of the lines other than $l$ reach all lines via $l$,
so only $\sum_{i=1}^{n-2}i$ deliveries are not made. The total number of
intersections is $\binom n2-\sum_{i=1}^{k-2}\binom{n_i}2$ if there are
$n_i$ lines with the $i$th slope other than the smallest or the largest
one. The maximum capacity of this RGMSN is
\[ \ge 1-\frac{\sum_{i=1}^{n-2}i}{\binom n2-\frac12(k-2)
\lceil\frac{n-2}{k-2}\rceil(\lceil\frac{n-2}{k-2}\rceil-1)}
=1-\frac{k-2}{(k-3)n}+O(n^{-2}).\]

When $k=3$, we have \[ \kappa_0=1-\frac{a+b-2}{2n}-
\frac{(n-a-b-1)(n-2)-(a+b)^2}{2n(ab+(a+b)(n-a-b))} \]
maximized either when $a=b<\frac14(\sqrt{5n^2-16n+12}-(n-2))$ or when
$b=1$ and $a>\frac12(\sqrt{5n^2-16n+12}-n)$. Both result in an upper bound
$1-\frac1{\sqrt n}+\frac{9}{8n}+O(\frac1{n\sqrt n})$ of the capacity.

Let $1\le a\le\frac{n-1}2$ be an integer. Consider the $b$ lines
$y=k$ ($0\le k<b$), the $a-1$ lines $y=x-k$ ($1\le k<a$), the $a$ lines
$y=-x+2b+a+k-2$ ($0\le k<a$), and the line $y=x-2(a+b-1)$ (see the figure
below). The capacity of this RGMSN is \begin{align*}f(a,b)
&=\frac{1}{n(a^2+2ab)}\bigg(\sum_{k=2}^{b+1}k+(b+1)\sum_{k=b+2}^{a+b+1}k
+(a-1)\sum_{k=b+2}^{a+b+1}k+a\sum_{k=1}^{a-1}k+b\sum_{k=a+b+2}^{2a+b}k\bigg)\\
&=\frac{a+b+1}{n}-\frac{n-1}{2n}\frac{b}{a^2+2ab}.\end{align*}
This attains the upper bound $1-n^{-1/2}+\frac{9}{8}n^{-1}+\Theta(n^{-3/2})$
above when $a$ is allowed to be non-integers. When $a$ is an integer,
the deviation from the upper bound is at most
\[ \max_{0\le t\le 1}\{f(a',b')-f(a'+t/2,b'-t)\}
=\frac{1}{2n\sqrt{n}}+O(n^{-2}), \] where
$f(a',b')=\max_{1\le a\le(n-1)/2} f(a,b)$,
so this construction attains the upper bound.

When $k=2$, if there are $a$ lines of one slope and $b$ lines of the
other slope, then the capacity is $\frac1{nab}\sum_{i=1}^a\sum_{j=1}^b(i+j)
=\frac12+\frac1n$.\qed

\begin{center}\begin{tikzpicture}[scale=0.5]
\draw[dashed](0,0)--(10,0);\draw[dashed](0,1)--(10,1);
\draw(0,-1)--(5,4);\draw(0,-2)--(6,4);\draw(6,-2)--(10,2);
\draw(3,4)--(8,-1);\draw(2,4)--(8,-2);\draw(2,3)--(7,-2);
\end{tikzpicture}\\Construction for three slopes with $a=3$, $b=2$
\end{center}

\begin{center}\bf4. Geometric realizability of combinatorial mobile
sensor networks\end{center}

By Theorems 3 and 4, only $O(\frac{\log n}n)$ of the RCMSN are realizable
as GMSN. Knuth proved that the number of GMSN of $n$ sensors is less
than $3^{\binom{n+1}2}$ [K, p.~39], but the number of RCMSN of $n$ sensors
is $\binom n2!$. And there is unlikely an efficient algorithm to determine
if an RCMSN is realizable as a GMSN.

\mn\textit{Proof of Theorem 5.} Shor showed that deciding
the stretchability of a pseudoline arrangement is NP-Hard [S].
Because all pseudoline arrangements in his paper contain only pseudolines
that intersect any additional vertical line at most once, we shall
consider only such pseudolines. We also assume that any additional
vertical line contains at most one intersection of the pseudolines.
Because of this, it is possible to draw two vertical lines such that
all intersections of the pseudolines fall between them, and each
pseudoline ``starts'' (at infinity) on the left of the two vertical lines
and ``ends'' on the right of the two vertical lines. 

Suppose the arrangement contains $n$~pseudolines, each two of them
intersect exactly once between the two vertical lines. We convert
this arrangement to a finite sequence of numbers as follows. Let all
the intersections ordered by their $x$-coordinate be $p_1$, \ldots,
$p_{\binom n2}$. For each $1\le k\le\binom n2$, we draw a vertical ray
downward from $p_k$ and let $a_k$ be the total number of intersections
this ray form with all the pseudolines. Then $a_k\in\{1,\ldots,n-1\}$,
and we have a sequence $\{a_k\}_{k=1}^{n(n-1)/2}$ (see the figure below). 

Let $\sigma$ be any permutation of $\{1,\ldots,n\}$. We can convert
the sequence $\{a_k\}$ to an RCMSN using the following algorithm: 

\beginalgo
   $\tau\=\sigma$
\n \b{for} $k$ \b{from} 1 \b{to} $\binom n2$
\n\_    $c_k\=\{\tau(a_k),\tau(a_k+1)\}$
\n\_    \b{swap} $\{\tau(a_k),\tau(a_k+1)\}$
\n \b{return} $\{c_k\}_{k=1}^{n(n-1)/2}$
\endalgo

Here $\sigma$ is the labels of the pseudolines at the vertical line
on the left of all intersections (ordered by their $y$-coordinates),
and $\tau$~stores the current labels of the pseudolines at an
additional vertical line immediately before and after each intersection,
such that the labels are consistent with $\sigma$ (guaranteed by line 4).
Now, the geometric realizabilities of all $n!$ RCMSNs that can
be generated from the pseudoline arrangement are equivalent by a relabeling
of lines using $\sigma$. Therefore, we need to prove that
the pseudoline arrangement (if exists) is uniquely determined from an RCMSN
up to a relabeling of lines and isomorphism (two arrangements are
isomorphic if the graphs, where the vertices are the regions in
the arrangement and the edges are adjacency of regions, are isomorphic). 

The following algorithm converts an RCMSN
$\{c_k=\{x_k,y_k\}\}_{k=1}^{n(n-1)/2}$
(where each $x_k,y_k\in\{1,\ldots,n\}$) into a sequence
$\{a_k\}_{k=1}^{n(n-1)/2}$ (where each $a_k\in\{1,\ldots,n-1\}$)
as described before, and also reports some clearly nonrealizable RCMSNs: 

\beginalgo
   $\tau\=\text{identity}\in S_n$ 
\n $T\=\text{linked list of } \{1,\ldots,n\} \text{ with no links}$
\n \b{for} $k$ \b{from} 1 \b{to} $\binom n2$
\n\_    \b{if} $T\text{.degree}(\tau(x_k))=2$ \b{or}
$T\text{.degree}(\tau(y_k))=2$
\n\_\_      \b{return} \emph{non-realizable}
\n\_    \b{else if not} $T$.linked$(\tau(x_k),\tau(y_k))$
\n\_\_      $T$.link$(\tau(x_k),\tau(y_k))$
\n\_\_      \b{swap} $\{\tau(x_k),\tau(y_k)\}$
\n \ $\sigma(1)\=T$.head
\n \b{for} $k$ \b{from} 2 \b{to} $n$
\n\_    $\sigma(k)\=T$.next$(\sigma(k-1))$
\n $\tau\=\sigma$
\n \b{for} $k$ \b{from} 1 \b{to} $\binom{n}{2}$
\n\_    \b{if} $|\tau^{-1}(x_k)-\tau^{-1}(y_k)|\ne1$
\n\_\_      \b{return} \emph{non-realizable}
\n\_    \b{else}
\n\_\_      $a_k\=\min\{\tau^{-1}(x_k),\tau^{-1}(y_k)\}$
\n\_\_      \b{swap} $\{\tau^{-1}(x_k),\tau^{-1}(y_k)\}$
\n \b{return} $\sigma$ \b{and} $\{a_k\}_{k=1}^{n(n-1)/2}$
\endalgo

In lines 1--8 we first determine which two numbers in the given RCMSN
might represent adjacent pseudolines on the left of both vertical lines
we inserted in a GMSN. $T$~is a list of consecutive lines, and
$\tau$~represents that ``the current $k$th pseudoline from the top is
the $\tau(k)$-th pseudoline on the left of both vertical lines
we inserted.'' Only adjacent pseudolines are allowed to intersect
because otherwise the pseudolines between them cannot extend across
the intersection. Thus, if we fail to build a whole list~$T$, no such
GMSN exists. On the other hand, if the list~$T$ is built,
it represents the relabeling of lines discussed in the previous
algorithm and is thus stored in~$\sigma$ (lines 9--11). In lines 12--18,
the meaning of~$\tau$ is the same as that in the previous algorithm,
and we reverse the previous algorithm to find $\{a_k\}$.
If we fail to reverse it, then at some point there must be two
nonadjacent lines that are required to intersect, so clearly
no such GMSN exists. 

From this algorithm, we see that if an RCMSN is geometrically
realizable, it corresponds to at most two pairs $(\sigma,\{a_k\})$
(because $T$ can be read from both sides), and their corresponding
pseudoline arrangements are isomorphic as they are mirror images of
each other. We shall ignore $\sigma$ because it does not change the
pseudoline arrangement. 

\begin{center}\begin{tikzpicture}
\draw(-1,1)--(2,1)--(4,3)--(5,3)--(6,4)--(7,4);
\draw(-1,2)--(0,2)--(2,4)--(5,4)--(6,3)--(7,3);
\draw(-1,3)--(0,3)--(1,2)--(2,2)--(3,1)--(4,1)--(5,2)--(7,2);
\draw(-1,4)--(1,4)--(2,3)--(3,3)--(5,1)--(7,1);
\draw(-1,0)--(7,0);\Label{right}{$\to$}{(6.5,-0.0215)};
\Label{right}{$x$}{(7,0)};\Label{left}{1}{(-1,1)};\Label{left}{2}{(-1,2)};
\Label{left}{3}{(-1,3)};\Label{left}{4}{(-1,4)};
\Label{below}{2}{(0.5,2.5)};\Label{below}{3}{(1.5,3.5)};
\Label{below}{1}{(2.5,1.5)};\Label{below}{2}{(3.5,2.5)};
\Label{below}{1}{(4.5,1.5)};\Label{below}{3}{(5.5,3.5)};
\Label{right}{(3)}{(0,3.5)};\Label{right}{(3)}{(3,3.5)};
\Label{right}{(3)}{(6,3.5)};\Label{right}{(2)}{(-1,2.5)};
\Label{right}{(2)}{(2,2.5)};\Label{left}{(2)}{(5,2.5)};
\Label{right}{(1)}{(1,1.5)};\Label{right}{(1)}{(3,1.5)};
\Label{right}{(1)}{(5,1.5)};
\Point{below}{0}{(0,0)};\Point{below}{1}{(1,0)};\Point{below}{2}{(2,0)};
\Point{below}{3}{(3,0)};\Point{below}{4}{(4,0)};\Point{below}{5}{(5,0)};
\Point{below}{6}{(6,0)};
\end{tikzpicture}\\Sequence: $(2,3,1,2,1,3)$. Colors in parentheses.
\end{center}

Now we need to prove that every sequence $\{a_k\}_{k=1}^{n(n-1)/2}$
where $a_k\in\{1,\ldots,n-1\}$ describes at most one pseudoline
arrangement. First, it describes one arrangement naturally:
Consider all points $(k,j)$ where $0\le k\le\binom n2$ and $1\le j\le n$
are integers. We draw a ray from each $(0,j)$ horizontally to the left
and from each $(\binom{n}{2},j)$ horizontally to the right. Then,
for each~$k$ we connect $(k-1,a_k)$ to $(k,a_k+1)$ by a segment,
$(k-1,a_k+1)$ to $(k,a_k)$ by a segment, and $(k-1,j)$ to $(k,j)$
by a segment for all $j\not\in\{a_k,a_k+1\}$ (see the figure above).
If this is not a pseudoline arrangement, then two curves must have
intersected twice, and the RCMSN corresponds to no pseudoline arrangement.
Otherwise, the RCMSN is generated by at least one pseudoline arrangement. 

Given a pseudoline arrangement whose sequence is~$\{a_k\}$,
we need to prove that it is isomorphic to the pseudoline arrangement above.
We give a region color~1 if it is adjacent to the region below all
pseudolines. Then we give a region color~$r$ if it is adjacent to an
already colored region with color~$r-1$ (see the figure above).
By definition, the number of
regions with color~$r$ is equal to the number of~$r$'s in~$\{a_k\}$ plus 1.
So in our two arrangements the number of regions with each color is equal.
Also, it is clear that in both arrangements, the leftmost and
rightmost regions with color~$r$ must be adjacent to the leftmost
and rightmost regions with color~$r\pm1$, respectively. If a region
immediately on the left of intersection~$p_j$ and a region immediately
on the right of intersection~$p_k$ have colors differing by one
(which is the only case they might be adjacent, by definition), then
they are adjacent if and only if $k<j$ because the pseudoline connecting
the two intersections is either a shared edge or an edge that separates
the two regions. There are no other possible cases of adjacent regions,
so the graph of adjacent regions is determined by the sequences $\{a_k\}$.
Therefore the two pseudoline arrangements we have are isomorphic. 

Hence, if we have an algorithm for the geometric realizability of RCMSNs,
then for every pseudoline arrangement we can convert it to an RCMSN
in polynomial time and determine if it is generated from a GMSN;
and the GMSN as a pseudoline arrangement must be isomorphic to
the given one because they correspond to the same RCMSN, as
discussed above.\qed

\mn\textit{Proof of Proposition 7.} First, we use the method in Theorem 6
to convert a CMSN into a pseudoline arrangement when possible.
If this is not possible, then the CMSN is clearly not realizable.
Now we construct a graph where the vertices represent the pseudolines,
and two vertices are connected by an edge if the two pseudolines
do not intersect. If any component in this graph is not a complete graph,
then the pseudoline arrangement is clearly not stretchable since
being parallel is transitive. If this graph has more than two
connected components, then the CMSN is also not realizable with at most
three slopes because we can find four pairwise intersecting pseudolines.

Without loss of generality, when the CMSN is realizable, we may assume
the slopes are 0 and 1. Let the lines be $y=a_1$, $y=a_2$, \ldots,
$y=x+b_1$, $y=x+b_2$, \ldots. The CMSN gives
a strict ordering of the $x$-coordinates of all intersections, and
it is realizable iff this linear program with strict inequalities
is feasible. This is solvable in polynomial time.\qed

\medskip If we regard two CMSN as equivalent if one is obtained by swapping
adjacent disjoint packets, then the CMSN is realizable if and only
if the pseudoline arrangement obtained from it is stretchable.
In this case, realizability as an RGMSN of three slopes becomes a linear
program as after an affine transformation, we may assume the slopes are
0 and $\pm1$. The complexity of geometric realizability into a fixed
$>3$ number of slopes is an open problem.

\begin{center}\bf Appendix. Maximum capacity of RCMSN\end{center}

\noindent\textbf{Theorem 8.} [G, Thm.~2.3]
The maximum possible capacity of an RCMSN or GMSN
with $n$~sensors is $1-\frac1n+\frac2{n^2}$.

\mn\textit{Proof.} Let the sequence of packets be $a_1,\ldots,a_N$
where $N=\binom n2$. If a graph is formed with vertices $\{1,\ldots,n\}$
$a_k,\ldots,a_N$, then the connected component containing $(a_k,b_k)$
has at most $N-k+1$ edges and thus at most $N-k+2$ vertices. Therefore,
$a_k$ can reach at most $N-k+2$ sensors for $k\ge N-n+2$, and
the total number of deliveries of a RCMSN or GMSN is no more than
\[ \sum_{k=1}^{N-n+2}n+\sum_{k=N-n+3}^N(N-k+2)=n(N-n+2)+\frac{n^2-n-2}2
=\frac{(n-1)(n^2-n+2)}2. \]
So the maximum capacity is at most \[ \frac{(n-1)(n^2-n+2)/2}{n^2(n-1)/2}=
1-\frac1n+\frac2{n^2}. \]
Now we place $n-2$ nonvertical and pairwise nonparallel lines randomly,
put a line with a slope less than the minimum slope of the previous
$n-2$ lines to the right of all previous intersections, and then
put a line with a slope greater than the maximum of
the previous $n-1$ lines to the right of all previous intersections.
Then every intersection of the first $n-1$ lines obviously can
reach all lines, and the last $n-1$~intersections can reach,
from left to right, $n,n-1,\ldots,2$ lines. This GMSN attains the upper
bound.\qed

\begin{center}\bf Acknowledgement\end{center}

This research was primarily done at my high school, Princeton
International School of Mathematics and Science. I thank Jesse Geneson
from Iowa State University for mentorship
and proposal of this project. I thank Qiusheng Li from my
high school for discussion about the problems. I thank Yongyi Chen
from MIT and Imre Leader from the University of Cambridge
for suggestions about the paper. I thank Tanya Khovanova, Pavel Etingof,
and Slava Gerovitch from the PRIMES-USA program in MIT
for the research opportunity.

\begin{center}\bf References\end{center}\parindent=0pt

\let\pbr=\par\def\par{\hangindent=2em\pbr}

[CRRZ] A.\ Casteigts, M.\ Raskin, M.\ Renken, and V.\ Zamaraev (2024),
Sharp Thresholds in Random Simple Temporal Graphs, \textsl{SIAM J.\ Comput.}
\textbf{53}:2, 346--388.

[G] J.\ Geneson (2018). Sharp extremal bounds on information
diffusion capacities of mobile sensor networks. Preprint.
\texttt{https://osf.io/n46qv}.

[GDGG] C.\ Gu, I.\ Downes, O.\ Gnawali, and L.\ Guibas (2018).
On the Ability of Mobile Sensor Network to Diffuse Information.
\textsl{Proc.\ 17th ACM/IEEE Int.\ Conf.\ Inf.\ Process.\ Sens.\ Netw.}
37--47.

[K] D.\ E.\ Knuth (1992), \textsl{Axioms and Hulls}.
Springer--Verlag, Berlin.

[S] P.\ W.\ Shor (1991). Stretchability of pseudolines is NP-hard,
In \textsl{Applied Geometry and Discrete Mathematics:
The Victor Klee Festschrift}, volume 4 of DIMACS Series in
Discrete Mathematics and Theoretical Computer Science, 
American Mathematical Society, pp.\ 531--554.

\end{document}